\documentclass[onecolumn]{webofc}

\usepackage[varg]{txfonts}   
\usepackage{lineno}
\newcommand{\hyt}{${}^3_\Lambda \rm{H}$ }
\newcommand{\hyh}{${}^4_\Lambda \rm{H}$ }

\begin{document}
\title{Hypernuclei and Antihypernuclei Production in Heavy-Ion Collisions}
%
%

\author{\firstname{Yue-Hang} \lastname{Leung}\inst{1}\fnsep\thanks{\email{yhleung@lbl.gov}} 
}

\institute{Nuclear Science Division, Lawrence Berkeley National Laboratory
          }

\abstract{%
In these proceedings, an overview of recent hypernuclei measurements in heavy ion collisions is presented. These results include the lifetime, $\Lambda$ binding energy, production yield, and directed flow in heavy ion collisions. These results provide constraints on the hyperon-nucleon interaction, enhance our understanding in hypernuclear structure, and provide insight on the production mechanisms of hypernuclei in heavy ion collisions.
}
\maketitle
\section{Introduction}
\label{intro}
Hypernuclei are nuclei containing at least one hyperon. Hypernuclei provide access to the hyperon-nucleon (Y-N) interaction, which is an important ingredient in the equation of state of high density nuclear matter, such as the core of neutron stars or the hadronic phase of a heavy ion collision. More than forty $\Lambda$-hypernuclei have been discovered, and through the systematic study of their $\Lambda$ separation energies ($B_{\Lambda}$), it can be inferred that the $\Lambda-N$ interaction is attractive, with a depth of approximately 30 MeV. More precise measurements on the lifetime, $B_{\Lambda}$, etc. can help us further constrain the $\Lambda-N$ interaction. On the other hand, $\Xi-N$, $\Sigma-N$ or $\Lambda-\Lambda$ interactions are not well understood due to the few double $s$ hypernuclei being discovered and studied.

Traditionally, hypernuclei are studied through techniques such as emulsion, $\gamma$-ray or strangeness exchange reactions. The first observations of hypernuclei and antihypernuclei from heavy ion collisions are from E864~\cite{Ref0} in 2004 and STAR~\cite{Ref1} in 2009 respectively. There are few advantages of studying hypernuclei in heavy ion collisions. Firstly, light nuclei are generally produced in copious amounts, thus enabling precise measurements on the hypernuclei structure, such as binding energy and lifetime. Also, the production mechanisms of light nuclei may be studied, providing input to the Y-N interaction as well as properties of the matter formed. Finally, it has been proposed that exotic hypernuclei may be produced in heavy ion collisions~\cite{Ref2}. Recent measurements of hypernuclei in heavy ion collisions provide new input to the Y-N interaction. 

\section{The Internal Structure of Hypernuclei}
\label{Sec:l}

\subsection{Hypertriton Lifetime Puzzle}
The hypertriton is a weakly bound object, leading to small overlap between the $\Lambda$ and the deuteron core. This leads to the theoretical expectation that the \hyt lifetime is close to the free $\Lambda$ lifetime $\tau_{\Lambda}$. However, STAR and HypHI has reported lifetimes $30-40\%$ shorter than $\tau_{\Lambda}$. This is commonly known as the hypertriton lifetime puzzle. 

\begin{figure}
\centering
\sidecaption
\includegraphics[width=7cm,clip]{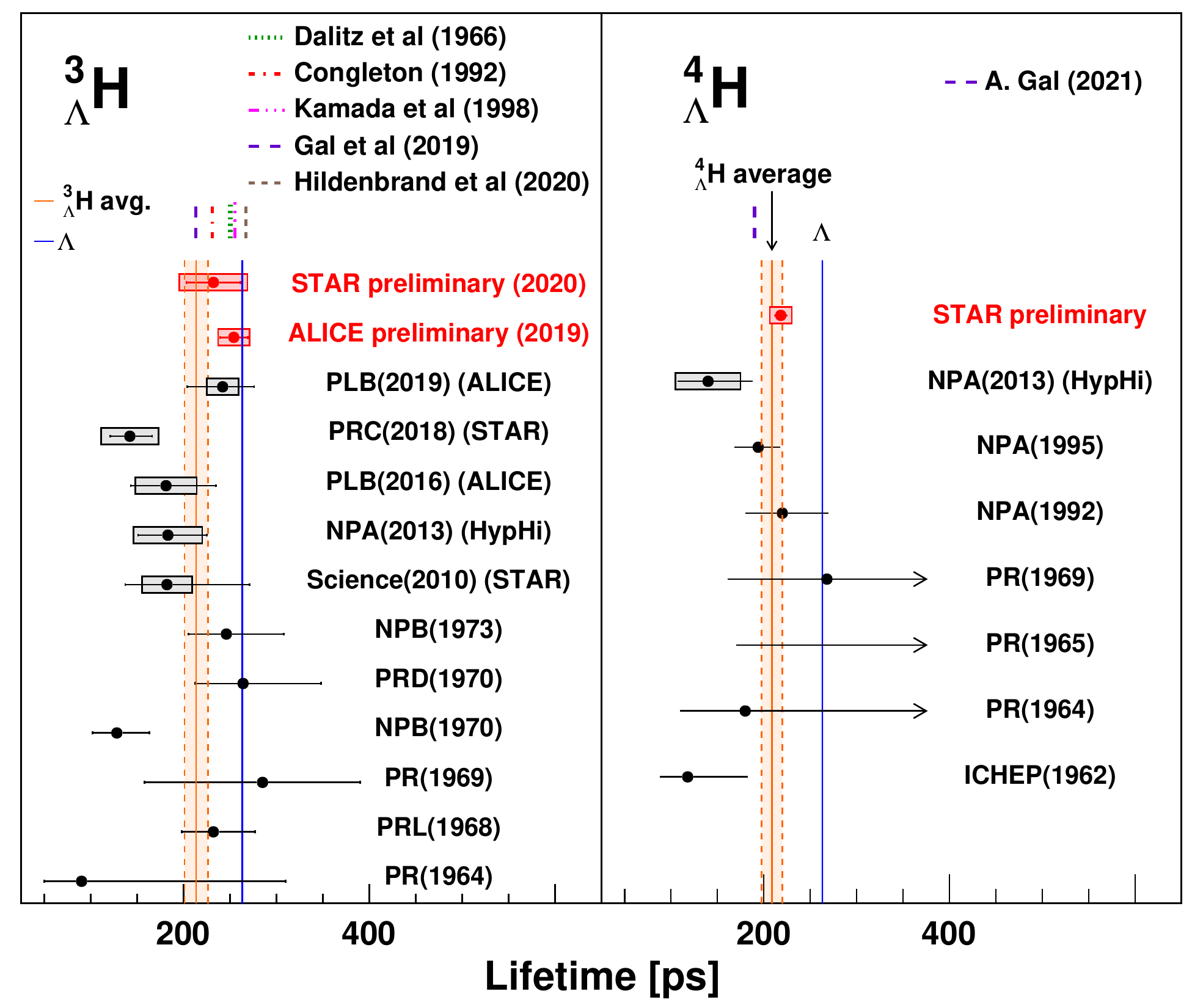}
\caption{\hyt (a) and \hyh (b) measured lifetimes, compared to previous measurements, theoretical calculations and the free $\Lambda$ lifetime.  The experimental average lifetimes and the corresponding  uncertainty  are  also  shown as orange bands.}
\label{fig-1}       
\end{figure}

Recently, STAR~\cite{Ref3} and ALICE~\cite{Ref4} have presented preliminary results of the \hyt lifetime. The results are shown in Fig.~\ref{fig-1}a. Taking into account of these results, the current experimental averaged \hyt lifetime is $81\pm5\%\tau_{\Lambda}$. Future measurements using high statistics data sets from STAR, ALICE and HADES~\cite{Ref5} are expected to further reduce the experimental uncertainty. On the theoretical side, recent studies from Gal et al.~\cite{Ref6} indicate that pion final state interactions are attractive and may decrease the \hyt lifetime by $\sim 10\%$. Predictions incorporating these considerations are consistent with the world average. In addition, the possible relationship between $B_{\Lambda}$ and lifetime has been studied in~\cite{Ref7,Ref8}. There is active progress both experimentally and theoretically towards a very precise understanding of the \hyt lifetime. 

\subsection{Hypertriton $B_{\Lambda}$}
The $\Lambda$ separation energy of the \hyt is defined as $B_{\Lambda}({}^{3}_{\Lambda}\rm{H}) = (m_{d}+m_{\Lambda}-m_{{}^{3}_{\Lambda}\rm{H}})c^{2}$, where $m_{d}, m_{\Lambda}$ and $m_{{}^{3}_{\Lambda}\rm{H}}$ are the deuteron mass, the $\Lambda$ mass and the \hyt mass. In 2019, STAR measured the $B_{\Lambda}$ of \hyt and ${}^{3}_{\bar{\Lambda}}\rm{\bar{H}}$ using data taken by the Heavy Flavor Tracker. The $B_{\Lambda}$s measured are consistent with each other, and the average is $0.41 \pm 0.12 (\rm stat.) \pm 0.11 (\rm syst.)$[MeV], seemingly in tension with
the result from emulsion experiments, $B_{\Lambda} = 0.13\pm0.05$ [MeV]. On the other hand, studies from~\cite{Ref9} showed that recalibrating the results from emulsion measurements gives an updated value of $B_{\Lambda} = 0.27\pm0.08$ [MeV], consistent with the STAR measurement. ALICE reported a precise new preliminary result of $B_{\Lambda} = 0.05 \pm 0.06 (\rm stat.) \pm 0.10 (\rm syst.)$[MeV]~\cite{Ref13}, further reducing the experimental uncertainty in  $B_{\Lambda}({}^{3}_{\Lambda}\rm{H})$. Precise measurements is important since it has a strong influence in the coalescence calculation of production yield in heavy ion collisions, as well as potential impact on lifetime calculations as well.
\begin{figure}[h!]
\centering
\sidecaption
\includegraphics[width=4.5cm,clip]{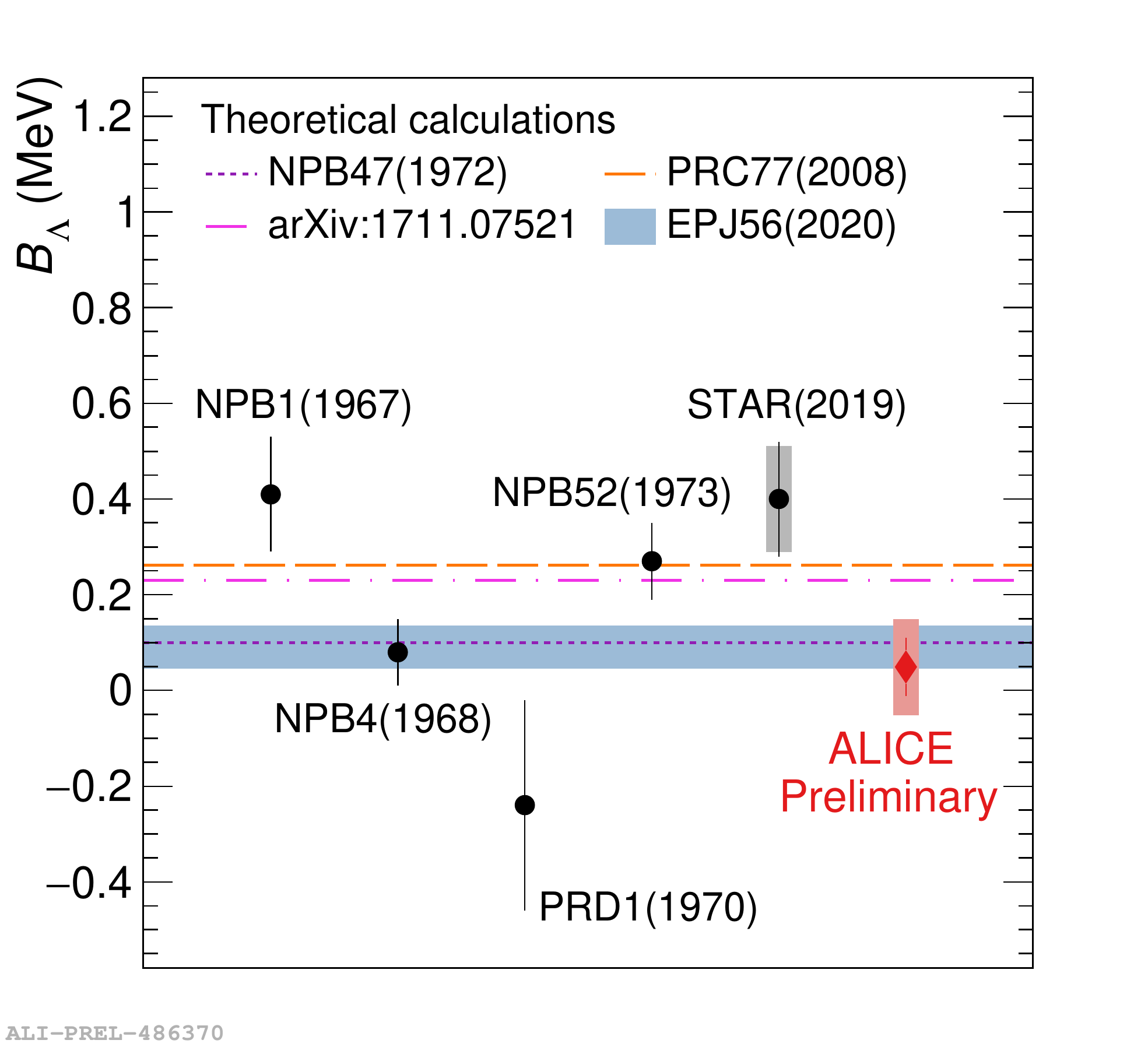}
\caption{Preliminary $B_{\Lambda}({}^{3}_{\Lambda}\rm{H})$ measurement from ALICE (red), compared with theoretical predictions and previous measurements. The values reported for previous measurements are recalibrated values reported in~\cite{Ref9}.}
\label{fig-3}       
\end{figure}

\subsection{Measurements of A=4 Hypernuclei}
\begin{figure}[h]
\centering
\sidecaption
\includegraphics[width=8cm,clip]{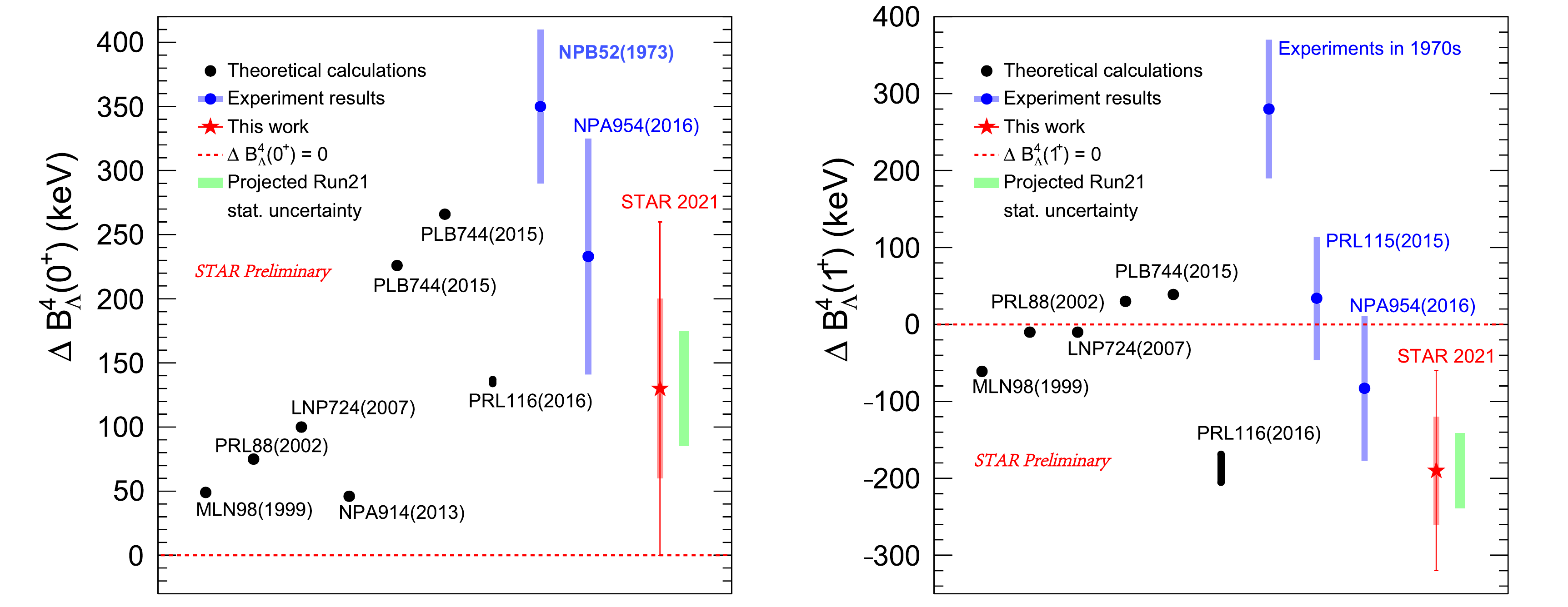}
\caption{The $B_{\Lambda}$ difference between \hyh and ${}^{4}_{\Lambda}\rm{He}$ in ground states (left) and in excited states (right) compared with theoretical model calculations (black dots) and previous measurements (blue dots). The green bands are projected statistical uncertainties from STAR run 2021 3 GeV data.}
\label{fig-2}       
\end{figure}

Besides \hyt, measurements on the lifetime and $B_{\Lambda}$ for $A=4$ hypernuclei have been reported by the STAR collaboration in this conference, using data from Au+Au collisions at $s_{\rm{NN}}=3$ GeV taken in 2018. The lifetime of \hyh is measured to be $217  \pm 8 (\rm stat.) \pm 12 (\rm syst.)$[ps], consistent with theoretical calculations from~\cite{Ref6}. The $B_{\Lambda}$ for \hyh and ${}^{4}_{\Lambda}\rm{He}$ have been measured. A non-zero $\Delta B_{\Lambda} = B_{\Lambda}({}^{4}_{\Lambda}\rm{He})-B_{\Lambda}({}^{4}_{\Lambda}\rm{H})$ is commonly known as charge symmetry breaking. The $B_{\Lambda}$ difference in the ground state is measured to be $130 \pm 130 (\rm stat.) \pm 70 (\rm syst.)$ keV, and is compared to theoretical calculations in Fig.~\ref{fig-2}. Combining the ground state result with transition energies measurements~\cite{Ref11, Ref12}, the $B_{\Lambda}$ difference in the excited state is also reported, and is consistent with model calculations. STAR detector upgrades and future data sets will reduce the statistical uncertainty, providing stronger constraints to theoretical calculations.

\section{Production of Hypernuclei in Heavy-Ion Collisions}
\label{Sec:lI}
In heavy ion collisions, light hypernuclei are expected to  be  abundantly  produced  at  low  energies due  to  the high  baryon  density. However,  the  production mechanisms of hypernuclei in heavy ion collisions are not well unuderstood, due to the scarcity of data.  Coalescence has been proposed as a possible production mechanism, particularly in central heavy ion collisions, while in peripheral collisions or small systems, production via the absorption of hyperons in the spectator fragments has been suggested. 

Hypernuclei yields have been measured by ALICE~\cite{Ref20}, STAR~\cite{Ref21}, and HypHI~\cite{Ref22} in the past.  The \hyt yield at mid-rapidity from Pb+Pb collisions at 2.76 TeV measured by ALICE are consistent with thermal model predictions~\cite{Ref23} and UrQMD~\cite{Ref24}. Multi-differential measurements in the production yield of hypernuclei are in different systems are essential in advancing our understanding in the field.

\subsection{${}^{3}_{\Lambda}\rm{H}$ and ${}^{4}_{\Lambda}\rm{H}$ Production from 3 GeV Au+Au Collisions at STAR}
\begin{figure}[h]
\centering
\sidecaption
\includegraphics[width=8cm,clip]{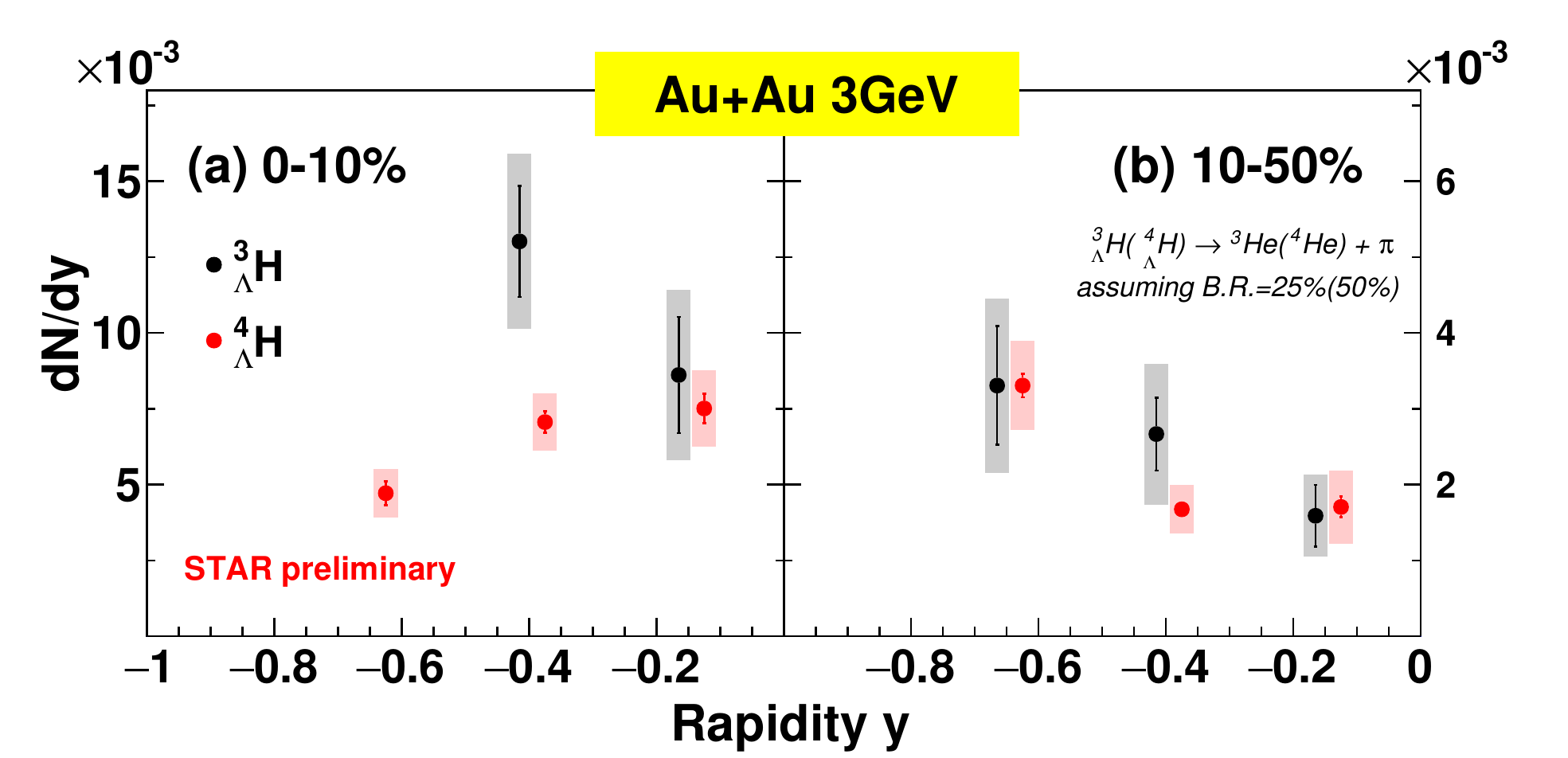}
\caption{$dN/dy$ as a function of rapidity $y$ for \hyt (black) and \hyh (red) for (a) $0-10\%$ centrality and (b) $10-50\%$ centrality Au+Au collisions at 3 GeV.}
\label{fig-8}       
\end{figure}

STAR has recently measured the \hyt and \hyh production yield at mid-rapidity in Au+Au collisions with $\sqrt{s_{\rm{NN}}}=3$ GeV, using data taken in 2018. The $p_{T}$ spectra are obtained, and the $p_{T}$ integrated yields are estimated by extrapolating to $0$ $p_{T}$. Systematic uncertainties are estimated by using different functional forms for extrapolation. The rapidity distributions of \hyt and \hyh are shown in Fig.~\ref{fig-8}. For \hyh, different trends in the rapidity distribution in central $(0-10\%)$ and mid-central $(10-50\%)$ collisions are observed. This is likely due to differences in collision geometry, possibly due to spectator reactions playing a larger role in non-central collisions.

\begin{figure}[h]
\centering
\includegraphics[width=10cm,clip]{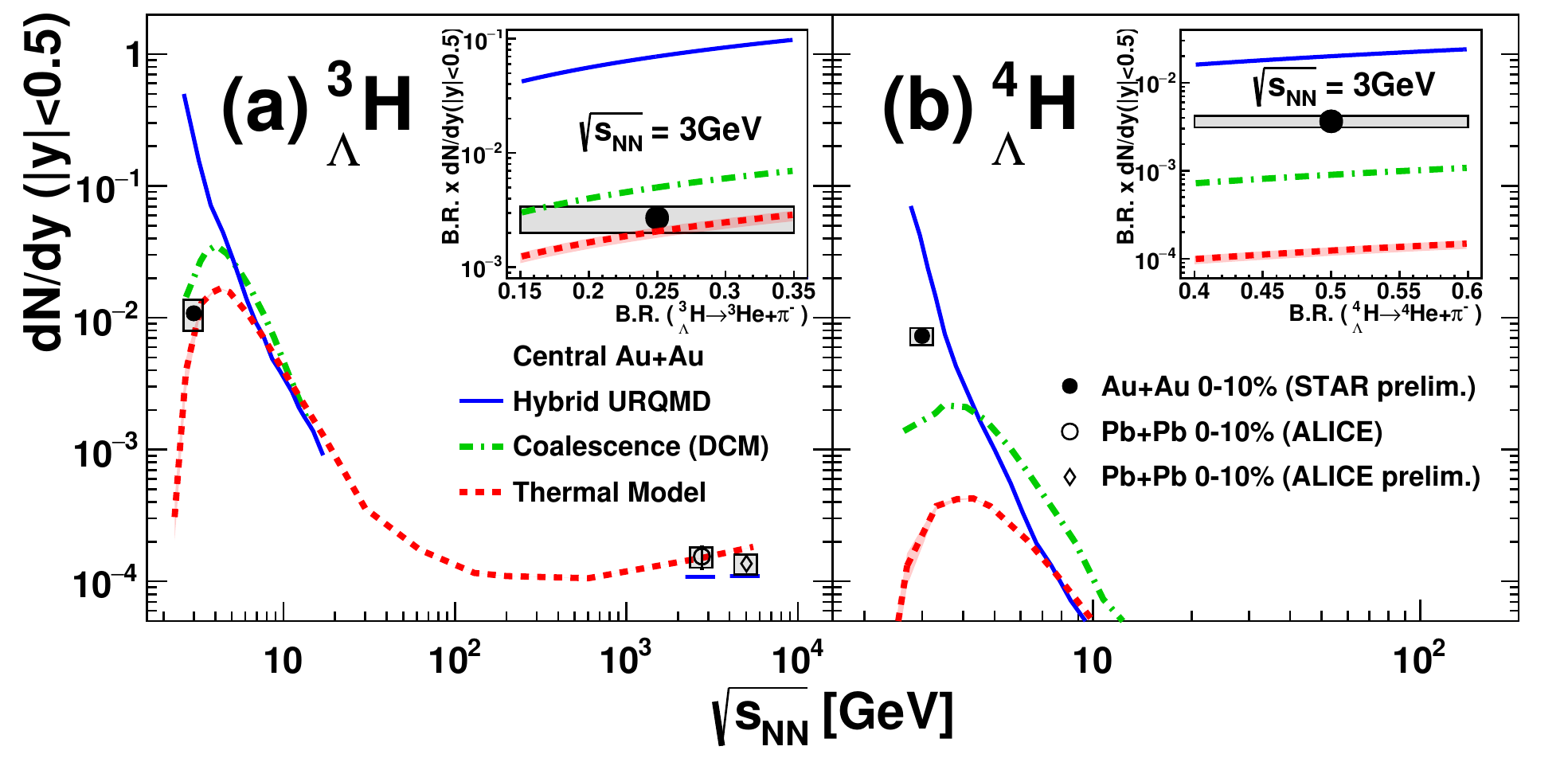}
\caption{(a)\hyt and (b)\hyh yields at $|y|<0.5$ as a function of beam energy in central heavy ion collisions. The symbols represent measurements while the lines represent different theoretical calculations. The data points assume a branching ratio of $25(50)$\% for $^{3}_{\Lambda}\rm{H}(^{4}_{\Lambda}\rm{H})\rightarrow ^{3}\rm{He}(^{4}\rm{He})+\pi^{-}$. The insets show the (a)\hyt and (b)\hyh yields at $|y|<0.5$ times the branching ratio as a function of the branching ratio.}
\label{fig-9}       
\end{figure}

The mid-rapidity $|y|<0.5$ yield in central heavy ion collisions is plotted as a function of beam energy in Fig.~\ref{fig-9}, and are compared to ALICE measurements at 2.76~\cite{Ref20} and 5.02 TeV~\cite{Ref4} Pb+Pb collisions, as well as theoretical calculations. The statistical thermal model~\cite{Ref23} which incorporates canonical suppression through the introduction of the canonical volume paramter $V_{C}$, can successfully describe the \hyt yield over few orders of magnitude of $\sqrt{s_{\rm{NN}}}$, while underestimating the \hyh yield at $3$ GeV. Similarly, the coalescence (DCM)~\cite{Ref24} also describes the \hyt yield at 3 GeV while underestimating \hyh. This discrepancy may possibly due to the coalescence parameters adopted in~\cite{Ref24}; the parameters for hyperons are assumed to be identical for nucleons and not constrained by data. The Hybrid URQMD model~\cite{Ref24}, although can describe the \hyt yield at LHC energies, overestimate the \hyt and \hyh yield at 3 GeV by an order of magnitude.

\begin{figure}[h]
\centering
\includegraphics[width=7.11cm,clip]{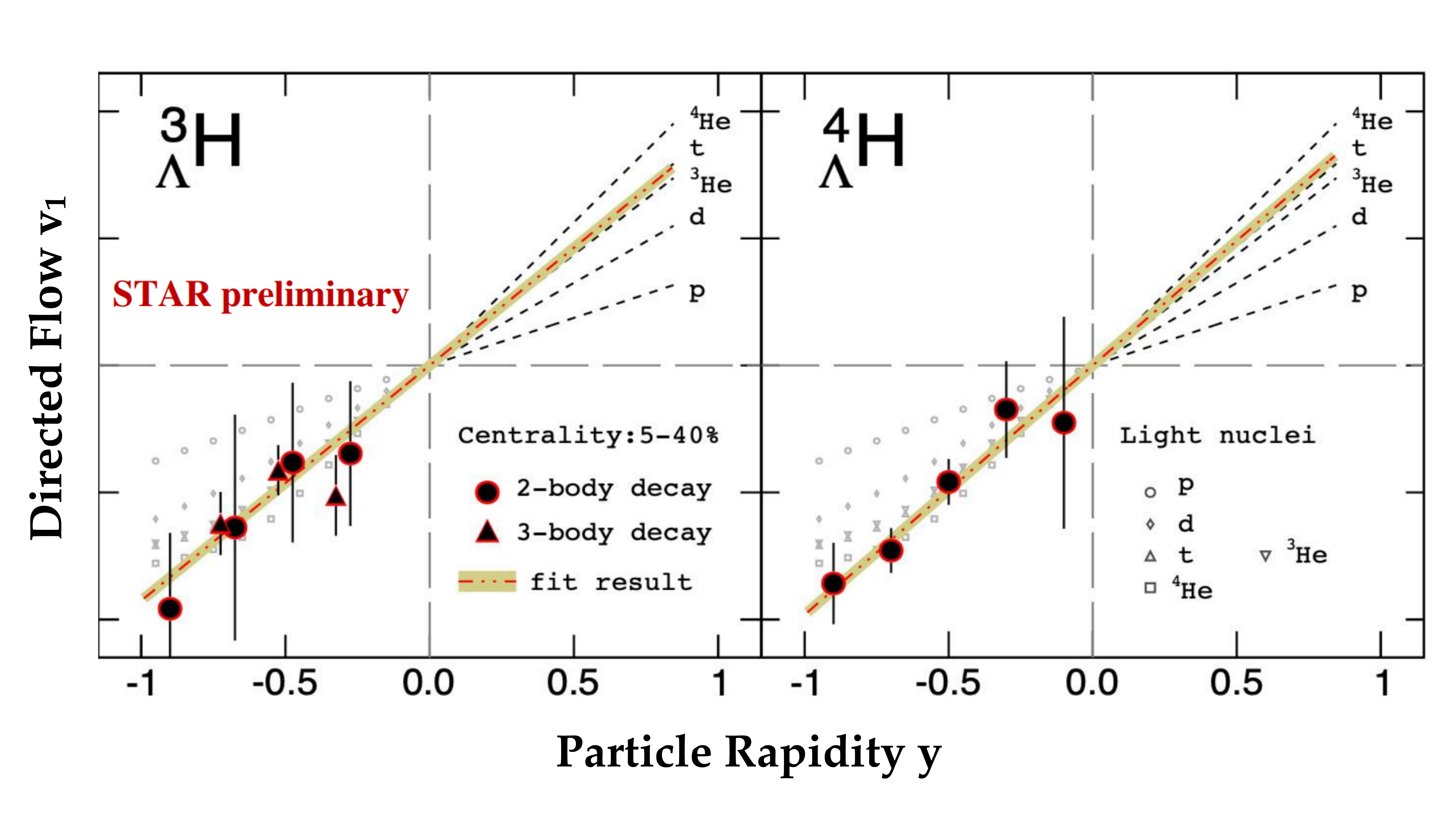}
\includegraphics[width=5.34cm,clip]{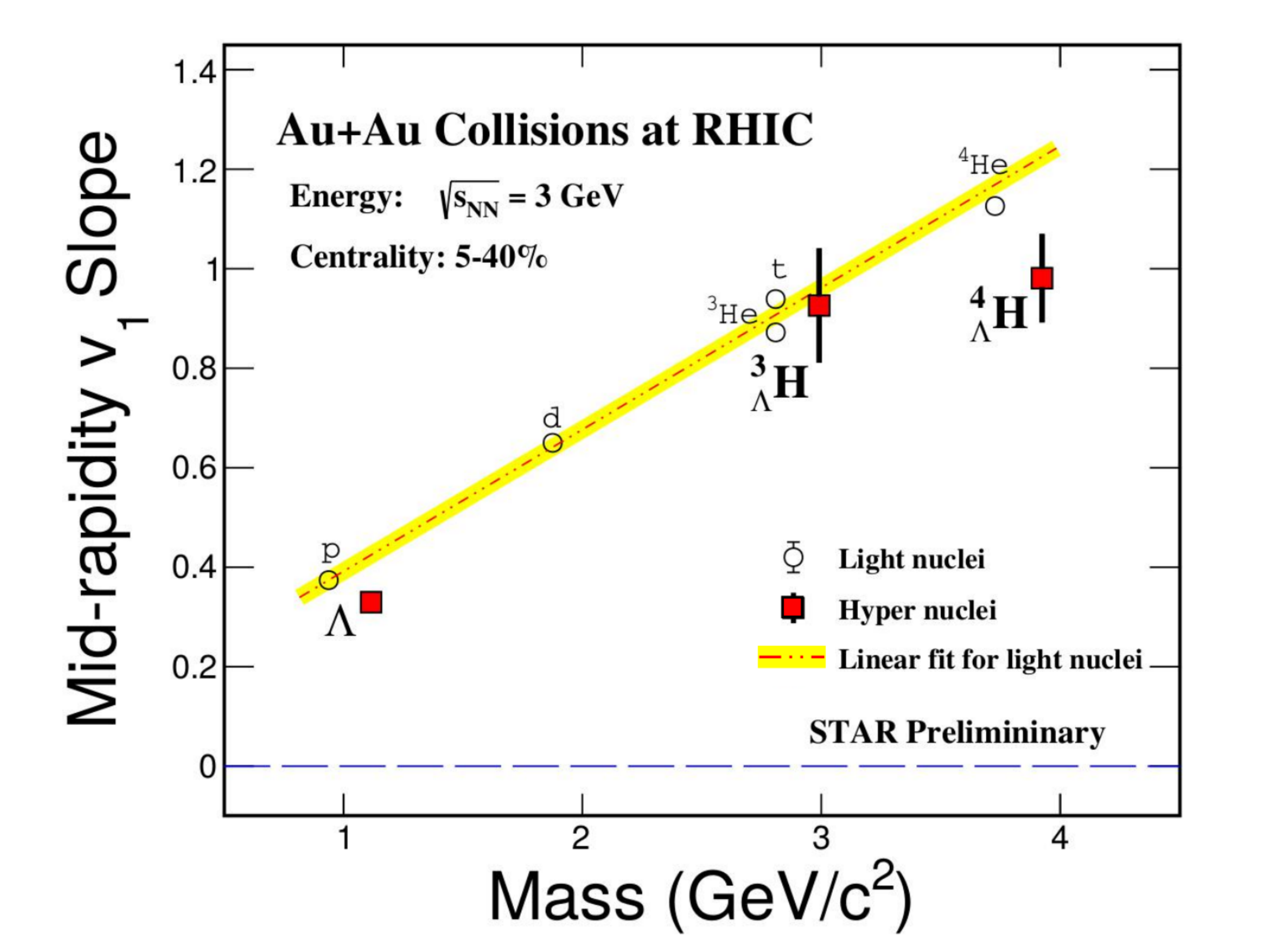}
\caption{(left)${}^{3}_{\Lambda}\rm{H}$ and \hyh $v_{1}$ as a function of rapidity. The red line represents a linear fit through the origin. (right) \hyt and \hyh mid-rapidity $v_{1}$ slope as a function of mass. The $dv_{1}/dy$ for light nuclei is shown for comparison.}
\label{fig-10}       
\end{figure}

Besides the production yield, the directed flow $v_{1}$ as a function of rapidity, for \hyt and \hyh in $5-40\%$ Au+Au collisions have been measured. The slope at mid-rapidity is obtained via a linear fit through the origin, and are shown in Fig.~\ref{fig-10}. The hypernuclei $dv_{1}/dy$ are compared to that of light hypernuclei. For light nuclei, a scaling behavior as a function of mass is observed, i.e. baryon number scaling. Since the $v_{1}$ slope of hypernuclei is similar to that of light nuclei with similar mass, the results are qualitatively consistent with hypernuclei production from coalescence of hyperons and nucleons at mid-rapidity. 

In summary, these results provide new quantitative input on the production mechanisms of loosely bound objects in high baryon density region.

\subsection{${}^{3}_{\Lambda}\rm{H}$ and ${}^{3}_{\bar{\Lambda}}\rm{\bar{H}}$ Production in Small Systems at ALICE}

\begin{figure}[h]
\centering
\includegraphics[width=5cm,clip]{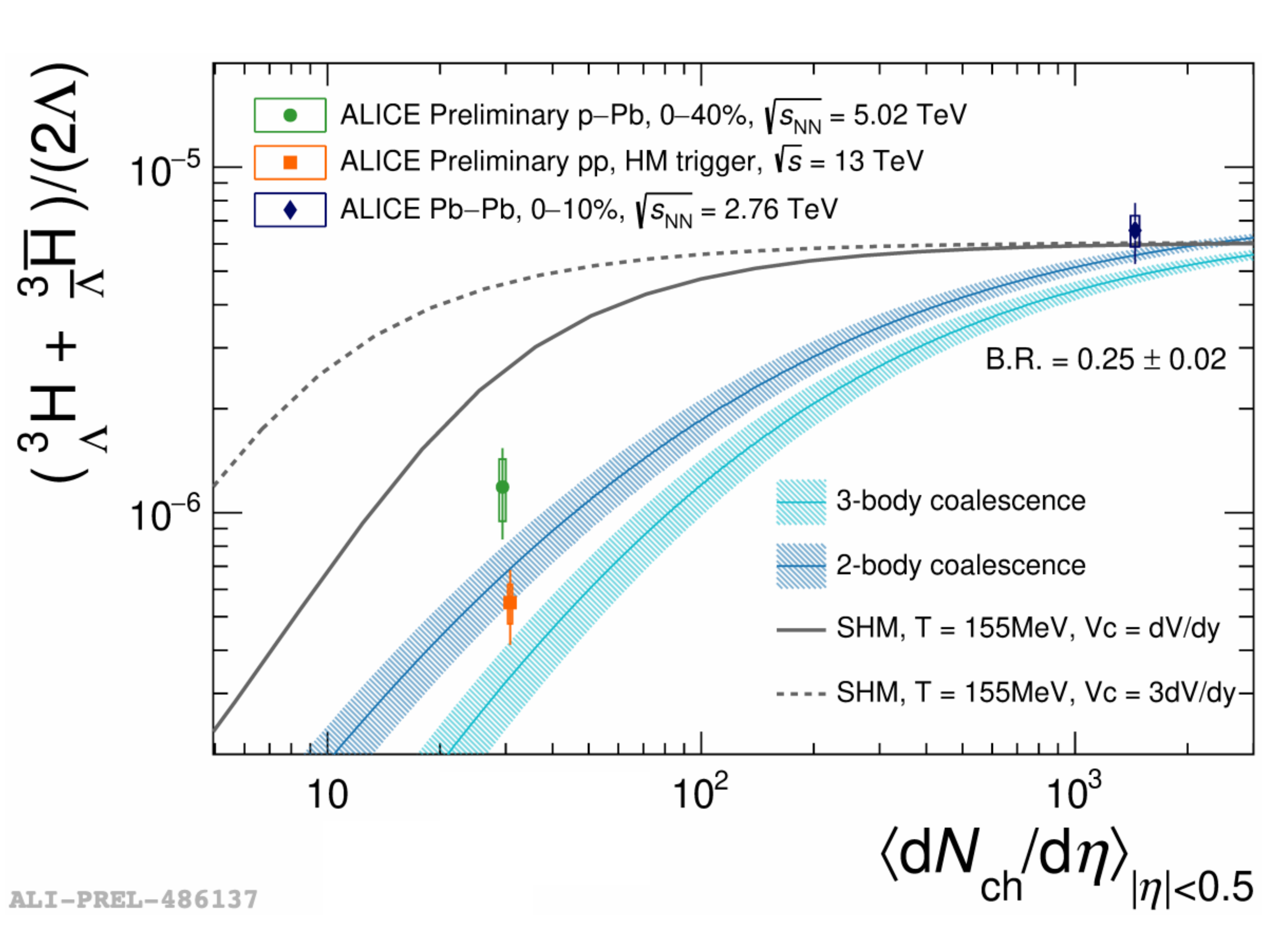}
\includegraphics[width=5cm,clip]{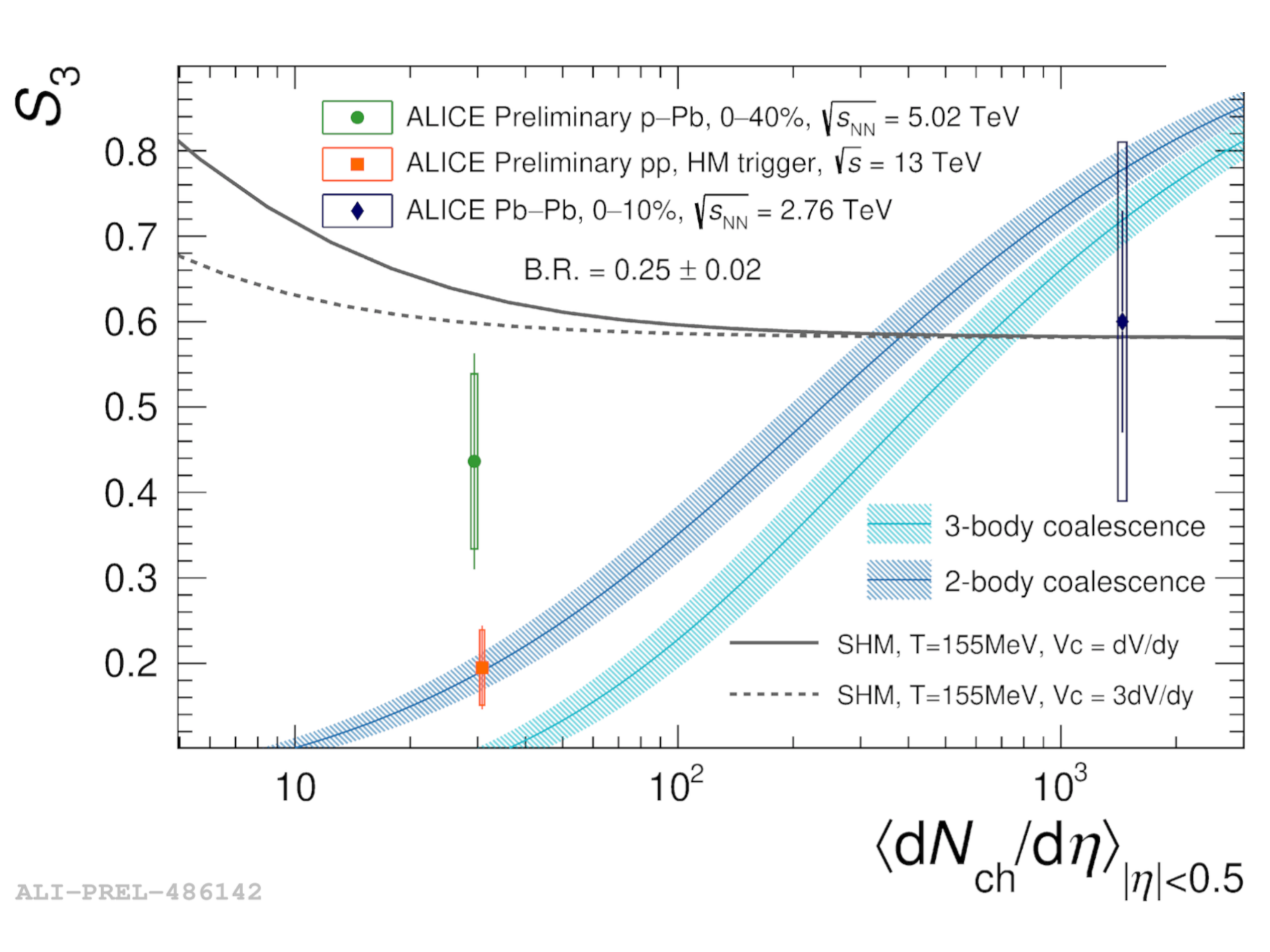}
\caption{${}^{3}_{\Lambda}\rm{H}$ (left) and $S_{3}$ (right) measurements in $p$+$p$ (orange), $p$+Pb (green) and Pb+Pb collisions(indigo) as a function of mean charged-particle multiplicity. The measured ratios are compared with expectations for the canonical statistical
hadronization~\cite{Ref15} and coalescence models~\cite{Ref14}.}
\label{fig-5}       
\end{figure}

\hyt and ${}^{3}_{\bar{\lambda}}\rm{\bar{H}}$ signals have been observed by ALICE in $0-40\%$ p+Pb collisions at $\sqrt{s_{\rm{NN}}}=5.02$ TeV and high multiplicity $p$+$p$ collisions at $\sqrt{s}=2.76$ TeV, both at $\sim 5\sigma$ significance~\cite{Ref13}. The ratio $({}^{3}_{\Lambda}\rm{H}+{}^{3}_{\bar{\Lambda}}\rm{\bar{H}})/2\Lambda$ and the strangeness population factor $S_{3}=({}^{3}_{\Lambda}\rm{H}/{}^{3}\rm{He})\times(p/\Lambda)$ are extracted and shown in Fig.~\ref{fig-5}. The $p$+$p$ data is consistent with 2-body coalescence~\cite{Ref14} and in tension with the statistical hadronization model (SHM)~\cite{Ref15}. For the $p$+Pb data, the measurement excludes certain configurations of the statistical model, while being consistent with 2-body coalescence. These new measurements give insight on the production mechanisms of hypernuclei in small systems, possibly with enough precision to distinguish between statistical hadronization and coalescence.

\section{Summary and Experimental Outlook}
\label{Sec:lII}

These proceedings present recent hypernuclei measurements from the STAR experiment and the ALICE experiment. The most precise \hyt and \hyh lifetime measurements to date, as well as measurements on $B_{\Lambda}$ of \hyt, \hyh and ${}^{4}_{\Lambda}\rm{He}$ are presented, providing strong constraints to theoretical models. The \hyt and \hyh rapidity distributions at $\sqrt{s_{NN}}=3$ GeV Au+Au collisions are presented. Thermal model and coalescence calcaulations are consistent with the \hyt yield, while no model can describe the \hyh yield. These new results provide insight to hypernuclei in production in the high baryon density region. Directed flow measurements are also presented, and the results are qualitatively consistent with expectations from production via coalescence. Finally, the \hyh yield is measured in $p$+$p$ collisions at 13 TeV and $p$+Pb collisions at 5.02 TeV. The results exclude certain configurations of the statistical hadronization model, and are consistent with coalescence models.

Future data from LHC can also shed light on the production mechanisms in smaller systems, while future data from STAR BES-II can help us understand the properties of high baryon density matter. Besides these current experiments, in the near future, data from heavy ion collisions at FAIR  (CBM, SIS), JPARC, NICA, HIAF, etc. are expected to contribute to our understanding of the Y-N and Y-Y interactions through the study of hypernuclei. These experiments cover a huge range in collision energy, allows the systematic study of hypernuclei production as a function of energy. Lifetime and binding energy of various hypernuclei can be measured to great precision, further enhancing our understanding of hypernuclear structure and the Y-N interaction. 

Similar to $\Lambda$ hypernuclei providing constraints to the Y-N interaction, the Y-Y interaction can be accessed through double-$\Lambda$ hypernuclei. Currently the $\Lambda-\Lambda$ potential is not well understood due to the lack of data. Only few double-$\Lambda$ hypernuclei have been discovered. According to thermal model predictions, the modest production rate of light double-$\Lambda$ hypernuclei, such as ${}^{5}_{\Lambda\Lambda}\rm{H}$, ${}^{5}_{\Lambda\Lambda}\rm{He}$ etc., gives discovery potential to heavy ion experiments. A discovery of these bound states, along with a measurement of the binding energy, are expected have a high impact on our understanding of the Y-Y interaction, which has important implications on nuclear and astrophysics.



%
%
%

\end{document}